%% file: main.tex
\documentclass[twocolumn,astrosymb]{aastex631}

\usepackage{graphicx}	% Including figure files
\usepackage{amsmath}	% Advanced maths commands
\usepackage{amssymb}	% Extra maths symbols
\usepackage{booktabs}
\usepackage{appendix}
\usepackage{comment}
\usepackage{enumitem}   % Itemlist formatting
\input{chocros}

\shorttitle{Radiative Flux from a Hot-Jupiter Simulation}
\shortauthors{Kafle, Cho \& Changeat}

\begin{document}

\title{Radiative Flux from a High-Resolution Atmospheric Dynamics 
Simulation of a Hot-Jupiter \\ for JWST and Ariel}

\author[0000-0003-2649-7978]{Jagat Kafle}
\affiliation{Martin A. Fisher School of Physics, Brandeis 
University, 415 South Street, Waltham, MA 02453, USA}
\correspondingauthor{Jagat Kafle}

\author[0000-0002-4525-5651]{James Y-K. Cho}
\affiliation{Martin A. Fisher School of Physics, Brandeis 
University, 415 South Street, Waltham, MA 02453, USA}

\author[0000-0001-6516-4493]{Quentin Changeat}
\affiliation{Kapteyn Institute, University of Groningen, 9747 AD, 
Groningen, NL}

\email{Emails:\ jagatkafle@brandeis.edu, jamescho@brandeis.edu, \\
q.changeat@rug.nl}

\begin{abstract}
    We present medium-wave ($\sim$0.5~$\mu$m to $\sim$13~$\mu$m) 
    radiative flux distributions and spectra derived from 
    high-resolution atmospheric dynamics simulations of an 
    exoplanet \WASPP.
    This planet serves to illustrate several important features.
    Assuming different chemical compositions for its atmosphere 
    (e.g., H$_2$/He only and $Z \in \{1, 12\}$ times solar 
    metallicity), the outgoing radiative flux is computed using 
    full radiative transfer that folds in the James Webb Space 
    Telescope (JWST) and Ariel instrument characteristics. 
    We find that the observed variability depends strongly on 
    the the assumed chemistry and the instrument wavelength 
    range, hence the probed altitude of the atmosphere.
    With H$_2$/He only, the flux and variability originate near 
    the 10$^5$~Pa level; with solar and higher metallicity, 
    $\sim$10$^3$~Pa level is probed, and the variability is 
    distinguishably reduced. 
    Our calculations show that JWST and Ariel have the 
    sensitivity to capture the atmospheric variability of 
    exoplanets like \WASPP, depending on the metallicity---both 
    in repeated eclipse and phase-curve observations.
\end{abstract}


 % Radiative Transfer

\section{Introduction}
Currently, there is a great need for rigorous estimates of exoplanet 
atmosphere variability.
The James Webb Space Telescope \citep[JWST;][]{Gardneretal06} now 
routinely observes the atmosphere of exoplanets (some only slightly 
larger than the Earth); and, the Ariel Telescope \citep{tinetti2021}, 
dedicated to 
observing thousands of exoplanet atmospheres, will soon concertedly 
characterize variability.
However, observational studies of large-scale weather patterns, which 
give rise to the variability, have remained limited thus far. 
This is largely due to the lack of repeated observations with 
signal-to-noise (S/N) that permits time-varying spectral features to 
be robustly identified; 
transit and eclipse observations, for example, frequently average the
data to boost the S/N, resulting in the loss of planet variability 
information \citep{Changeatetal24}. 
Even when the S/N is adequate, observations are typically not repeated, 
due to the observing time constraints on highly oversubscribed 
facilities.
Observations with the Hubble Space Telescope (HST) as well as the 
Spitzer and Kepler telescopes \citep[e.g.,][]{spitzer2007,kepler2014}, 
prior to the JWST, are often combined with observations from various 
JWST instruments, and at different epochs, to increase the wavelength 
coverage and characterize other properties of the planets. 

On the modeling side, high-resolution hydrodynamics simulations have 
consistently shown  dynamic, complex temperature and tracer 
distributions in hot-Jupiter atmospheres \citep[e.g.,][]{Choetal03, Choetal08,Choetal15,SkinCho21,Choetal21,Skinetal23}. 
In these simulations, giant storms and large-amplitude waves induce 
quasi-periodic temperature flux signatures on the large scale by 
transporting and mixing patches of hot as well as cold air. 
The spatiotemporal variability has been initially predicted and 
called to attention by \citet{Choetal03}, who suggested that such 
variability could be detected in observations:
for example, motion-induced changes in the temperature field would 
lead to observable variations in the spectra of the planetary 
atmospheres. 
In addition to opening a new window to weather and climate studies, 
identifying variability in the spectra would concomitantly help 
constrain the exoplanet dynamics models themselves.

In this paper, we focus on the exoplanet \WASPP.
This is a particularly interesting target for study. 
It orbits a F-type star WASP-121 and is an ``ultra-hot'' giant planet 
with an equilibrium temperature of $\sim$2360~K \citep{Delretal16}.
It has been observed multiple times. 
For example, it has been observed four times with the HST Wide field 
Camera 3 Grism 141 (WFC3-G141): one transit in June 2016, one eclipse 
in November 2016, and two phase curves in March 2018 and February 2019. 
Previous studies with HST, TESS, Spitzer, JWST, and ground-based 
facilities have revealed the distribution of water vapor, hydrogen ions 
(H$^{-}$), radiative absorbers (VO and TiO) as well as other atomic 
species (Ba, Ca, Cr, Fe, H, K, Li, Mg, Na, V, Sr), indicating complex 
chemical 
processes \citep[see, e.g.,][and references therein]{Changeatetal24}. 

\citet{Changeatetal24} have recently shown atmospheric variability of 
\WASPP\ by combining HST observations with high-resolution dynamics 
simulations.
Here we extend that work in two ways: 1)~we assess the pressure level 
(altitude) probed by thermal emission observations 
%%%(i.e., eclipse and phase-curves) 
and 2)~we assess the atmospheric variability in mid-infrared 
observations, when accurate dynamics simulations are utilized to 
obtain planetary fluxes and spectra. 
Importantly, the simulations are performed at very high, numerically
converged resolution and use a forcing setup informed by careful 
atmospheric retrievals from HST 
observations \citep{Changeatetal22,Edw_2023}.
Hence, the simulations are arguably the most realistic representations 
of the ultra-hot-Jupiters' flow and temperature distributions to date. 
The observables (e.g., spectroscopic thermal flux) so obtained are 
presented here to help guide future observation strategies with
next-generation telescopes, such as the JWST and Ariel, as well as 
to delineate the conditions under which the flux cannot be assumed to 
originate mostly near the $10^5$~Pa pressure level.

\section{Methodology}\label{sec:methodology}

In this study, we carefully post-process 8.5 contiguous planet days of 
global dynamics simulation outputs.
The dynamics and radiative transfer~(RT) are not coupled; the coupled 
study will be presented elsewhere. 
The duration is long enough to contain a full variation cycle, and the 
variation within is typical over the entire duration of the simulation.
The main steps in our methodology is summarized below.
For full descriptions of the dynamics model, simulations, RT, and 
chemistry, we refer the reader to \citet{SkinCho21} and 
\citet{Changeatetal24}. 

\subsection{Atmospheric Dynamics Simulation}

The dynamics simulations are performed with the parallel pseudospectral 
code, BoB \citep{Scotetal04,Polietal14}, which solves the 
three-dimensional (3D) traditional primitive equations in the pressure 
($p$) coordinate at high resolution.
Note that by high resolution it is meant a resolution of T341L50---i.e., 
degree and order of 341 each in the Legendre expansion of the field 
variables and 50 $p$ levels---employed here; 
the numerical algorithm implemented (spectral plus 16$^{\rm th}$-order hyperviscosity) generates fields which are comparable to those of at 
least $2000\! \times\! 1000$ horizontal grid resolution in a finite 
difference simulation \citep[e.g.,][]{SkinCho21}.
This is because of the exponential convergence property of the spectral algorithm: each doubling of the spectral resolution increases the 
accuracy by five to ten fold over doubling of the number of grid points 
in the conventional grid methods \citep[see, e.g.,][]{Boyd00,ThraCho11}.  
The resolution ensures that the governing equations are accurately solved
and fast, small-scale phenomena are captured \citep{Choetal21,SkinCho25}. 

\begin{figure*}
    \centering    
    \includegraphics[width=0.40\textwidth]{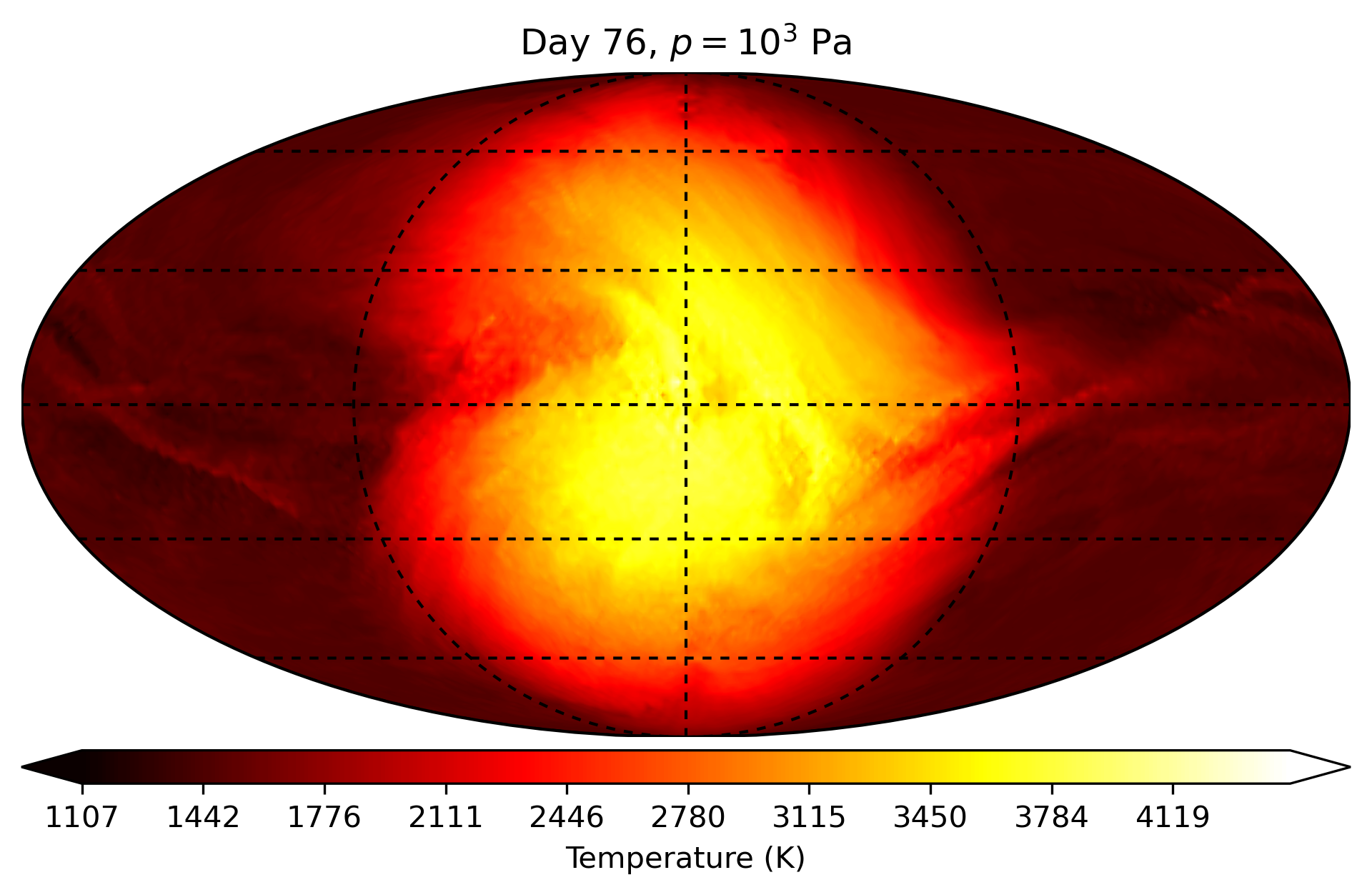}
    \hspace{1.6cm}
    \includegraphics[width=0.40\textwidth]{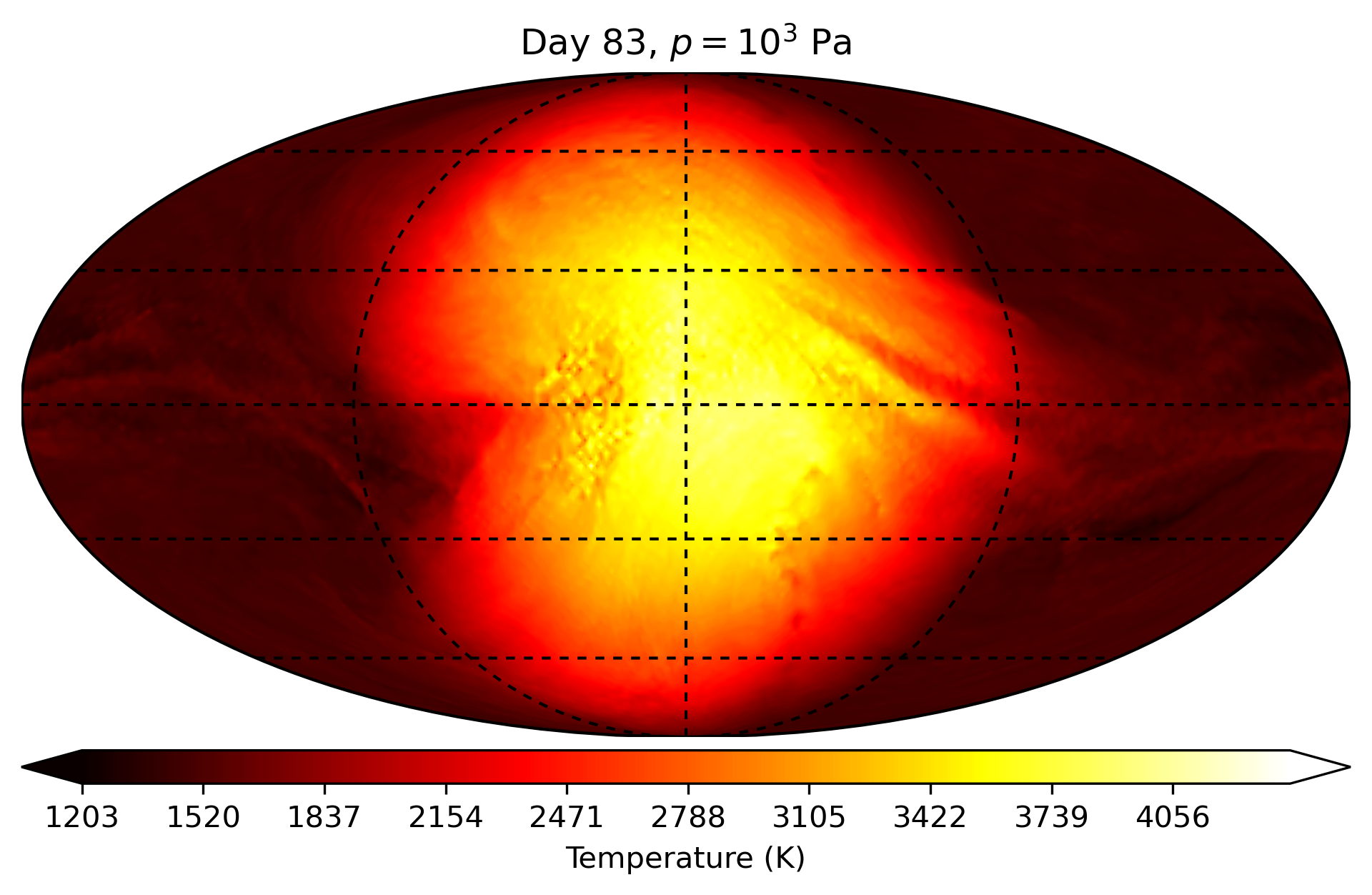}\\*[0.3cm]
    \includegraphics[width=0.40\textwidth]{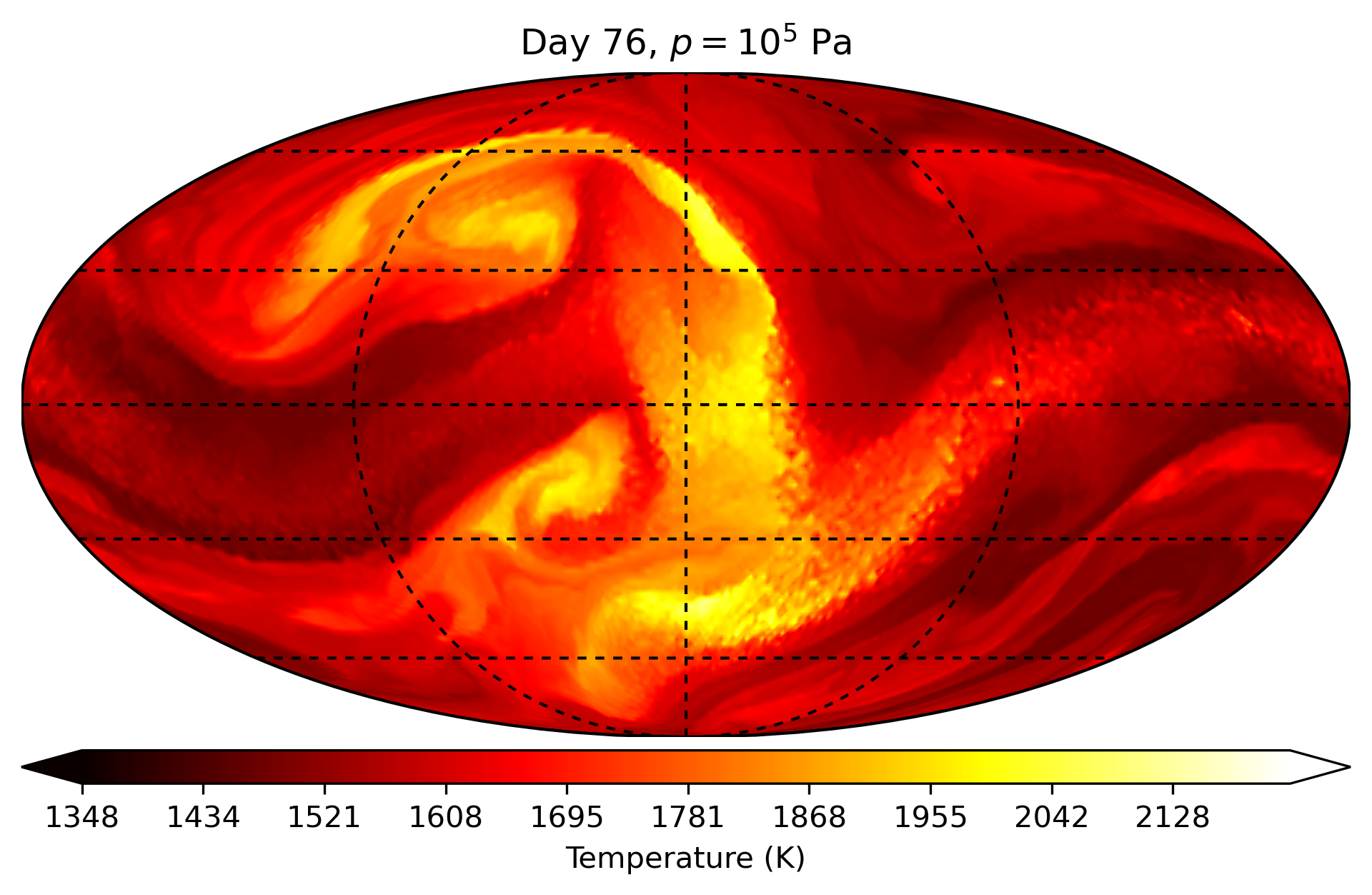}
    \hspace{1.6cm}
    \includegraphics[width=0.40\textwidth]{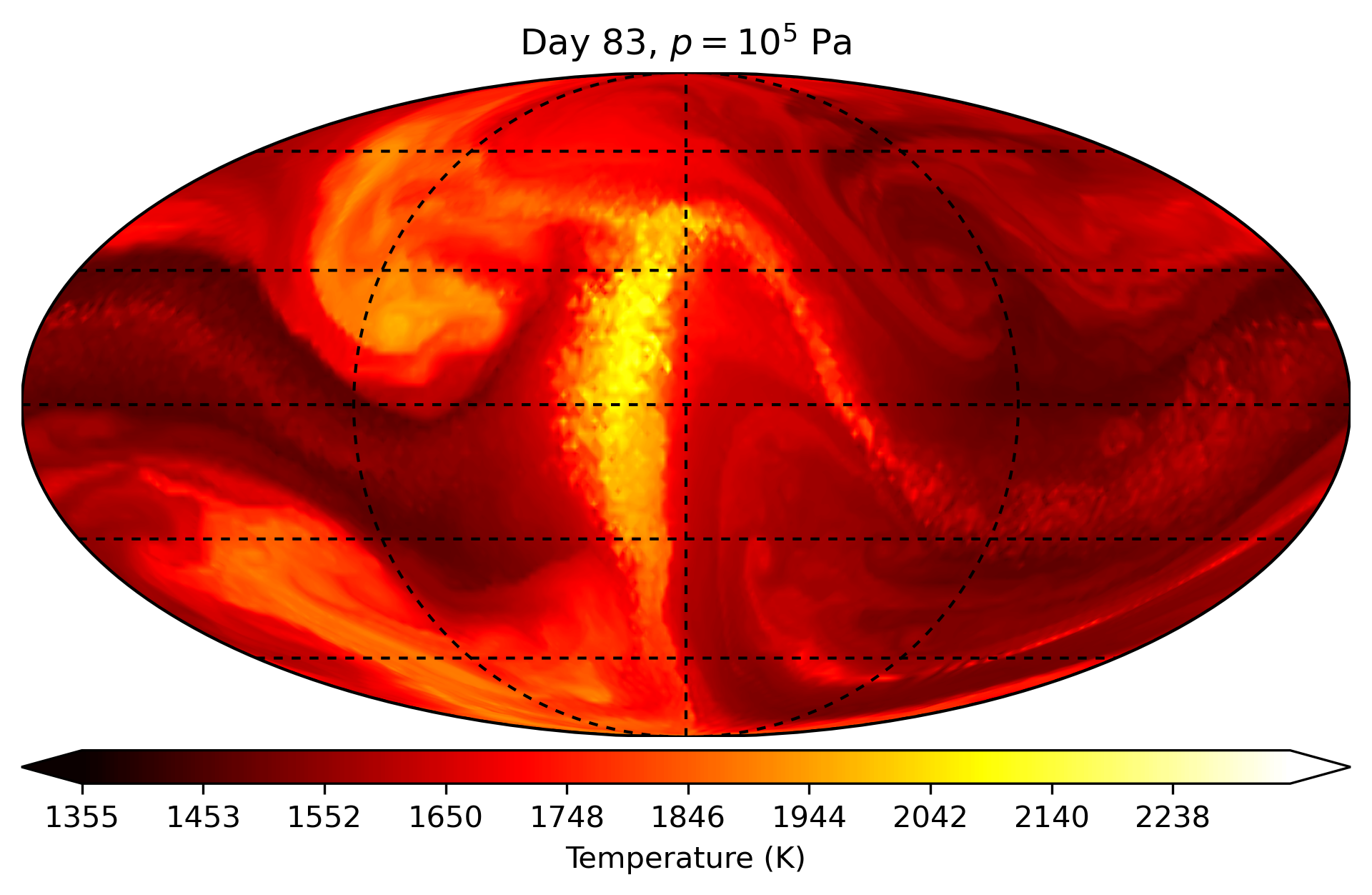}
    \caption{Temperature field $T(\varphi,\vartheta)$ at day $t = 76$ 
    (left column) and day $t = 83$ (right column) at the $p = 10^3$~Pa 
    (top row) and $p = 10^5$~Pa (bottom row) pressure levels, in 
    Mollweide projection centered on the substellar point;
    the two days shown are near the beginning and end of the 
    post-processed calculations. 
    On the large scale, the temperature distributions at the two times 
    are similar at the upper level (top row) but very dissimilar at the 
    lower level (bottom row);
    note that the grossly similar distribution at the upper level is 
    different on the small scale and varies in time.
    Temperature fields such as these from high-resolution dynamics 
    simulations are post-processed to obtain accurate chemical species 
    distributions and medium-wave outgoing thermal fluxes.}
    \label{fig:temp-map}
\end{figure*}

\subsection{Radiative Transfer Post-processing}

The temperature fields $T(\varphi,\vartheta,p,t)$, where $\varphi$ is 
the longitude and $\vartheta$ is the latitude, obtained from dynamics 
simulations (see, e.g., Figure \ref{fig:temp-map}) are post-processed 
using three different chemistry assumptions:
\begin{itemize}[leftmargin=\dimexpr 26pt-.3cm]
    \item[1.] {\it No active chemistry}\,:\ the atmosphere is composed 
    only of H$_2$ and He, with the main absorption coming from the 
    continuum (Collision Induced Absorption, CIA).

    \item[2.] {\it Solar metallicity}\,:\ the atmosphere has solar 
    abundances of the main elements ($Z = 1$), and the chemistry is 
    modeled using GGChem \citep{ggchem2018}.

    \item[3.] {\it Enhanced metallicity}\,:\ the atmosphere is 
    enriched, similar to that of Jupiter's atmosphere, and the 
    metallicity in GGChem is set to twelve times the solar value 
    ($Z = 12$).
\end{itemize}
The above assumptions are used to evaluate the impact of composition 
and chemistry on the $p$ level probed by observations---hence the 
observed variability, as will be seen below. 
After the chemistry of the main molecules $\{{\rm H}_2, {\rm He}, 
{\rm H}_2{\rm O}, {\rm CO}, {\rm CO}_2, {\rm CH}_4, {\rm TiO}, {\rm VO}, 
{\rm FeH}\}$ is obtained from GGChem (see Figure~\ref{fig:chemistry}, 
Appendix for a sample species distributions), the atmosphere is 
post-processed using the one-dimensional~(1D) plane-parallel RT scheme 
in TauREx3 \citep{Al-Refaie_2021, Al-Refaie_2022}. 

Full RT calculation is performed for each $(\varphi, \vartheta)$ point 
of the simulation output, taking into account vertical temperature 
distribution and chemical composition as well as the atmospheric path 
length and viewing angle. 
The RT scheme includes the absorption from all the relevant 
species---H$_2$O \citep{polyansky_h2o}, CH$_4$ \citep{ExoMol_CH4_new}, 
CO \citep{li_co_2015}, CO$_2$ \citep{Yurchenko_2020}, 
TiO \citep{mckemmish2019}, VO \citep{McKemmish_2016_vo}, and FeH \citep{Bernath_2020_FeH}---using ExoMol 
line-lists \citep{exomol2018, Chubb_2021_exomol, Tennyson_2024} at 
$\mathcal{R} = 15,000$ resolution. 
We also include CIA for the H$_2$--H$_2$ and H$_2$--He 
pairs \citep{abel_h2-h2, abel_h2-he} as well as Rayleigh Scattering \citep{cox_allen_rayleigh}. 
The simulation outputs are processed at 6 hour intervals, enabling a 
smooth time-evolving spectra of the thermal emission to be constructed. 
For select days, we also obtain the phase-dependent planetary emission 
seen from different viewing angles, since it is useful for JWST and 
Ariel observations.

\begin{figure*}
    \centering
    \includegraphics[width=0.49\textwidth,height=0.40\textwidth]{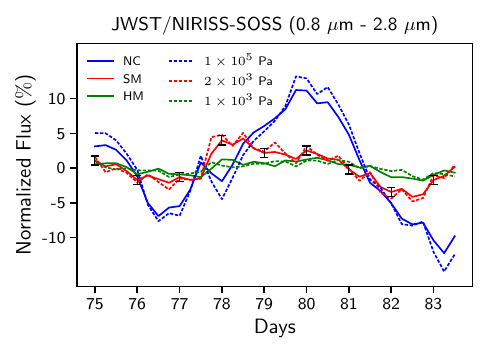}
    \includegraphics[width=0.49\textwidth,height=0.40\textwidth]{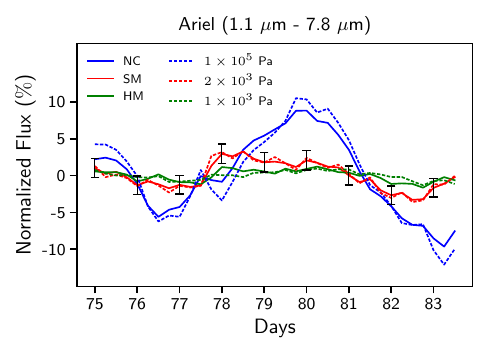}
    \caption{Normalized flux $\scrF$ in secondary eclipse (i.e., 
    dayside) as function of time, integrated over wavelength 
    ranges of JWST/NIRISS-SOSS (left) and Ariel (right); the 
    normalization is such that 
    $\scrF = (\calF\, /\, \overline{\calF}) - 1$, where 
    $\overline{\calF}$ is the mean flux over the duration shown. 
    Full radiative transfer (RT) fluxes (full lines) and blackbody 
    brightness temperature fluxes (dashed lines) at 
    $p \in \{1 \times 10^5,\, 2 \times 10^3,\, 1 \times 10^3\}$~Pa 
    levels are shown.
    In addition, atmospheres with three different compositions 
    (and chemistry) are shown: 
    H$_2$/He atmosphere dominated by collision-induced absorption 
    (NC), solar metallicity atmosphere (SM), and super-solar 
    metallicity atmosphere~(HM). 
    In the high pressure regions, higher amplitude variability is 
    observed for NC atmospheres; 
    assumed composition and chemistry change the variability, 
    with high metallicity atmosphere generally leading to reduced 
    variability. 
    Note also that the simple blackbody brightness and the full RT 
    calculations match very closely, provided that the brightness
    calculation is performed at the $p$ level probed by the 
    observation. 
    JWST and Ariel uncertainties for \WASPP\ secondary eclipse are 
    also displayed and show that both telescopes would be able to 
    observe the variability of this planet, depending on the 
    metallicity.}
    \label{fig:ar_jwst_flux}
\end{figure*}

\subsection{Brightness Temperature Post-processing}

Here, each $(\varphi, \vartheta)$ point is assumed to emit a 
blackbody radiation. 
With the $T(p)$ distribution at each point, the  spectral radiance 
is computed using the Planck distribution. 
The spectral radiance is disk-integrated over $\varphi$ and 
$\vartheta$, weighted by a cosine projection (to the surface normal) factor, and over the ranges of wavelength~$\lambda$ covered by JWST 
and Ariel, to obtain the flux at each $p$-level.\footnote{Note that different normalizations for the flux are used in this paper, to 
highlight different aspects; 
the normalizations do not qualitatively affect the basic results 
presented, but are nonetheless distinguished as needed for 
completeness and clarity.}
The ranges for JWST/NIRISS-SOSS, Ariel, and JWST/NIRSpec-G395H are 
$\lambda \in [0.8, 2.8]~\mu$m, $\lambda \in [1.1, 7.8]~\mu$m, and 
$\lambda \in [2.8, 5.2]~\mu$m, respectively (see 
Figure~\ref{fig:ar_jwst_flux}, and also 
Figure~\ref{fig:jwst_nirspec} in Appendix).

\subsection{Instrument Simulation}\label{sec:instrument-sim}

The post-processed outgoing flux obtained using the full RT and 
the brightness temperature methods is convolved with an instrument 
model (e.g., for JWST and Ariel) to ascertain observational 
performances. 
For JWST, we use the instrument simulator, ExoCTK 
Pandexo \citep{Batalha_2017}.
We simulate the spectra for the NIRISS-SOSS and NIRSpec-G395H 
instruments using the recommended setup (i.e., 636 integrations 
with 5 groups for NIRISS and 664 integrations with 34 groups for 
NIRSpec), since similar observations have been approved in previous 
JWST Cycles \citep{2017jwst.prop.1201L, 2021jwst.prop.1729M}. 
For Ariel, we utilize the official radiometric model 
ArielRad \citep{Mugnai_2020, Mugnai_2022} at Tier 3 resolution 
(Code versions: ArielRad v2.4.26, ExoRad v2.1.111, Payload v0.0.17) to estimate the performances of NIRSpec and AIRS. 
We use the standardized eclipse observation setup for Ariel as 
described in \citet{Mugnai_2020} and  \citet{tinetti2021}, 
which corresponds to the same in- and out-of-transit durations 
(i.e., 2.9\,h) as in the JWST simulations.

\section{Results}\label{sec:results}

Figure~\ref{fig:ar_jwst_flux} presents the main results of this 
paper.
The wavelength-integrated, normalized flux $\scrF(t)$ from planet 
days $t \in [75.0, 83.5]$ are shown for atmospheres with three 
different chemical compositions (NC, SM, and HM).
The starting time and duration are chosen because the variability 
exhibited is typical and because a full variation cycle is 
captured clearly.
The fluxes are combined with instrument models covering three 
$\lambda$ ranges, $[0.8,2.8]~\mu$m, $[1.1, 7.8]~\mu$m, and 
$[2.8, 5.2]~\mu$m, for JWST/NIRISS-SOSS, Ariel, and 
JWST/NIRSpec-G395H, respectively.
The first two suffice to illustrate the main points of our 
results; 
the latter provides additional details, and is therefore 
included in the Appendix.
The error bars in Figure \ref{fig:ar_jwst_flux}, obtained as 
described in Section~2.4, show the uncertainties for $\scrF(t)$ 
that would be observed by JWST/NIRISS-SOSS (0.64\%) and Ariel 
(1.28\%) at secondary eclipse. 
The dynamics-derived RT calculations (full lines) are shown for 
atmospheres with H$_2$ and He (only), solar metallicity 
($Z = 1$), and super-solar metallicity ($Z = 12$) compositions 
(labeled NC, SM, and HM, respectively, in the figure).
Also shown are brightness temperature fluxes (dashed lines), 
integrated over a disk centered at the substellar point over 
the appropriate $\lambda$ ranges, at the indicated $p$ levels; 
the levels are those from which most of the flux originates in 
NC, SM, and HM atmospheres.
Several features can be readily seen in the figure.

\subsection{Radiative Transfer and Simple Brightness Fluxes}
\label{sec:rtvsbrigh}

Firstly, the $\scrF$ at secondary eclipse is highly variable in time, 
but the amplitude of variability is dependent on the composition and 
chemistry---both as expected. 
The amplitude is high for the H$_2$/He atmosphere, with most of the 
flux contribution coming from $p \sim 1 \times 10^5$~Pa level.
In contrast, although the variability clearly present, its amplitude is 
reduced (\,$\lesssim$ 7\% peak-to-peak) for the $Z = 1$ and $Z = 12$ 
atmospheres, with the majority of the flux coming from lower $p$ levels 
($\sim$$2 \times 10^3$~Pa and $\sim$$1 \times 10^3$~Pa, respectively).
This is due to the greater opacity in these atmospheres and to the 
shorter thermal relaxation (i.e., ``radiative cooling'') timescales 
at the lower $p$ levels.
The latter leads to a nearly stationary (on the large scale), hot patch 
of atmosphere near the substellar point (see Figure~\ref{fig:temp-map}, 
top row); 
the variability is markedly reduced in repeated secondary eclipses 
because the {\it sampling} is effectively ``in phase'' with the 
spatial variation of the $T$ field.
The microstructure in the variability indicates that the hot patch in 
fact is not exactly stationary and the flow and $T$ fields are 
weakly baroclinic (vertically slanted); that is, there is a fluctuation 
over the reduced variability and the SM and HM fluxes are slightly
out of phase (in time) with the NC flux.

Note that the simple blackbody brightness temperature flux is, in 
general, a very good proxy for the full RT-derived flux---provided 
that the fluxes are computed at the $p$ level probed by observations; 
compare the dashed lines with the full lines for the three atmospheres.
This is wholly consistent with the argument forwarded in many past 
dynamics studies \citep[e.g.,][]{Choetal03,Choetal21,SkinCho22,Skinetal23,SkinCho25}:
when the $p$ level at which the flux emerges is known, the blackbody 
brightness temperature flux is sufficient for assessing the variability 
(at least for a \WASPP-like planet atmosphere).  
We remind the reader here, however, that the dynamics, chemistry, 
and RT are not coupled in these simulations.
A coupled simulation is likely to show quantitatively different 
variability patterns over space and time.
Results from dynamics--RT coupled simulations at high-resolution will 
be presented elsewhere.

\subsection{Spectral Dependence of Flux Variability}\label{sec:spectr}

Secondly, because JWST/NIRISS-SOSS, Ariel, and JWST/NIRSpec-G395H cover 
different $\lambda$ ranges, the variability observed by them is also 
different. 
For example, in Figure \ref{fig:ar_jwst_flux}, JWST/NIRISS-SOSS and 
Ariel show peak-to-peak variability of $\sim$25\% and $\sim$20\%, 
respectively, for the NC atmosphere. 
For JWST/NIRSpec-G395H, peak-to-peak variability dramatically reduces 
to approximately half that of Ariel (Figure \ref{fig:jwst_nirspec}, 
Appendix), despite the higher S/N from its larger collecting area. 
Nevertheless, our results indicate that both JWST and Ariel are 
sensitive enough to capture the variability as well as to delineate 
reduced variability---due to, e.g., super-solar metallicity.

\begin{figure}%%%[htb!]
    \centering
    \includegraphics[width=\linewidth,height=0.90\linewidth]
    {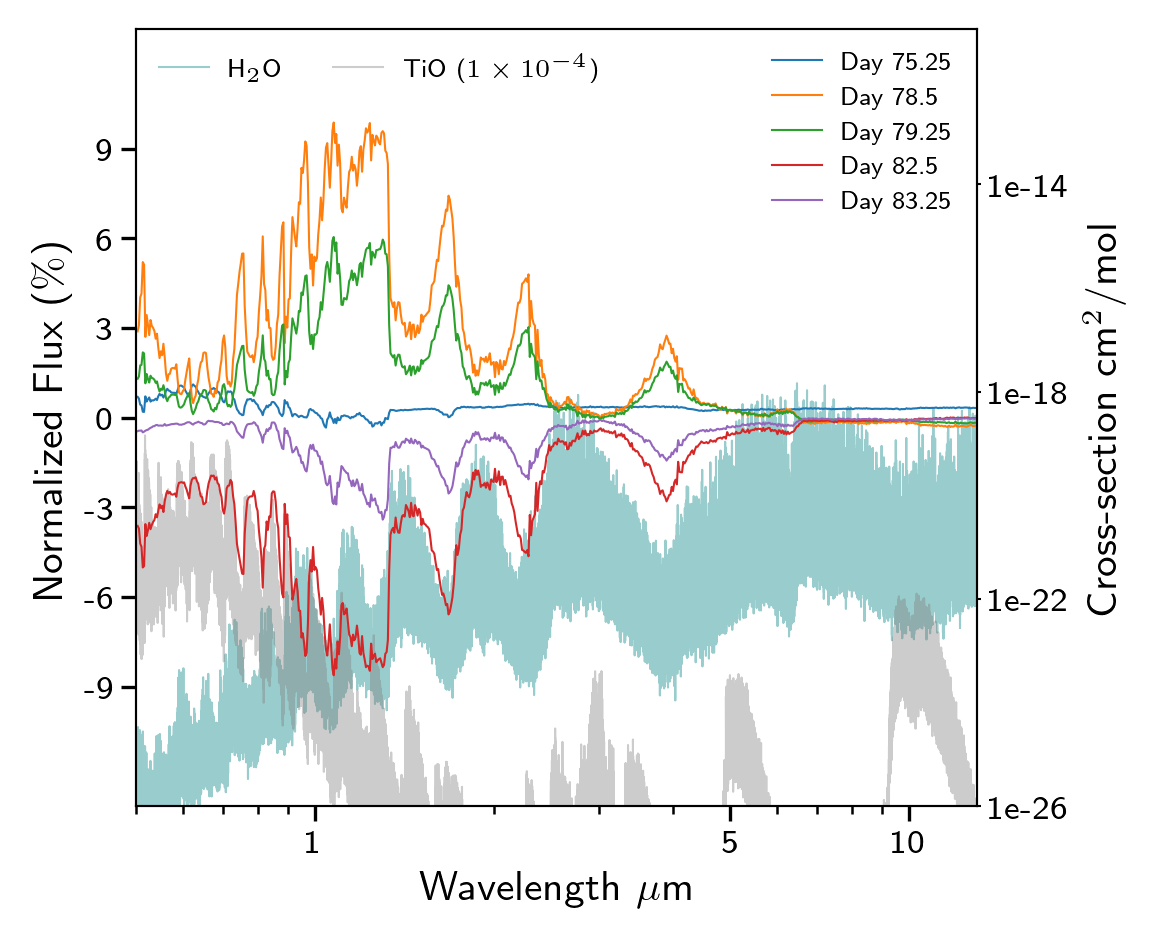}
    \caption{Normalized flux $\scrF_\lambda$ as a function of 
    wavelength  $\lambda$ using full RT for the $Z = 1$ atmosphere; 
    $\scrF_\lambda$ is obtained by integrating over a disk centered 
    on the substellar point (i.e., secondary eclipse), at the times 
    indicated. 
    The $\scrF_\lambda$ spectrum is sensitively dependent on 
    $\lambda$ (hence its coverage).
    In the range, $\lambda \in [1, 5]~\mu$m, the peaks and troughs 
    in $\scrF$ overlap with the troughs in the H$_2$O cross-section 
    (shaded cyan area). 
    In this atmosphere, H$_2$O is the main opacity source; in its 
    absorption windows, the outgoing flux originates from lower 
    pressure levels, which have diminished variability amplitudes. 
    The variability changes at different days, as the chemical 
    composition changes on the planet due to atmospheric motion.}
    \label{fig:flux_variation}
\end{figure}

The above is shown more explicitly in Figure \ref{fig:flux_variation}. 
It presents the normalized flux spectrum, 
$\scrF_\lambda = \scrF_\lambda(\lambda,t)$, over 
$\lambda \in [0.5, 13]~\mu$m for the $Z = 1$ atmosphere at secondary
eclipse. 
Displayed are spectra at various times, including those with maximum 
variability from the mean over days $\overline{\scrF_\lambda}$. 
As seen in the figure, there is a strong $\lambda$ dependence, with 
maximum peak-to-peak variability reaching up to $\sim$18\% at 
$\lambda \sim 1\, \mu$m. 
In contrast, the variability at longer wavelengths 
($\lambda \gtrsim 5 \, \mu$m) is much smaller at $\sim$1\%. 
Thus, as expected, the $p$ level probed is different, depending on 
the $\lambda$ range (or instrument) considered. 
The variability that would be seen by JWST/NIRISS-SOSS, Ariel, and JWST/NIRSpec-395H in Figures~\ref{fig:ar_jwst_flux} and 
\ref{fig:jwst_nirspec} is different because NIRSpec-G395H covers 
much redder wavelengths ($\,\gtrsim 2.8~\mu$m), where less variability 
is seen (Figure \ref{fig:flux_variation}). 
We note that Ariel's $\lambda$ range covers both the highly variable 
and not variable regions of the $\scrF_\lambda$ spectrum.

In Figure \ref{fig:flux_variation}, we also overlay the H$_2$O 
and TiO cross-sections, in order to illustrate the correlation of $\scrF_\lambda$ with these main opacity sources for the $Z = 1$ 
atmosphere. 
We observe large variability at the troughs in the H$_2$O 
cross-section (e.g., at 
$\lambda \sim \{1.1,\, 1.3,\, 1.6,\, 2.3,\, 4.0\}~\mu$m). 
At these wavelengths, H$_2$O does not absorb much.
Thus, the observation probes higher $p$ (lower altitude) levels, 
which are much more variable than at lower $p$ (higher altitude) 
levels; see Figure~\ref{fig:temp-map}. 
Similar characteristics can be seen for TiO at shorter wavelengths 
(e.g., $\lambda \sim \{0.5,\, 1.2\}~\mu$m).
Note that the variation also depends on the chemistry, as shown in 
Figure \ref{fig:ar_jwst_flux}.
The situation is complex and difficult to model reliably: here more 
observations would help to better constrain the atmospheric models.

\subsection{Phase-Dependence of Flux Variability}\label{sec:phase}

Thus far, we have discussed the flux variability when the planet 
is always observed in secondary eclipse.
However, simulations show moving thermal structures (i.e., large 
and/or intensely hot or cold regions) that could be captured in 
additional modes of observation---e.g., phase curves. 
Identifying such structures provides more stringent constraints 
on the atmospheric dynamics and its modeling. 
Hence, studying the planet's emission at different viewing 
angles is useful.  

\begin{figure}%%%[htb!]
    \centering
    \includegraphics[width=\linewidth,height=1.3\linewidth]{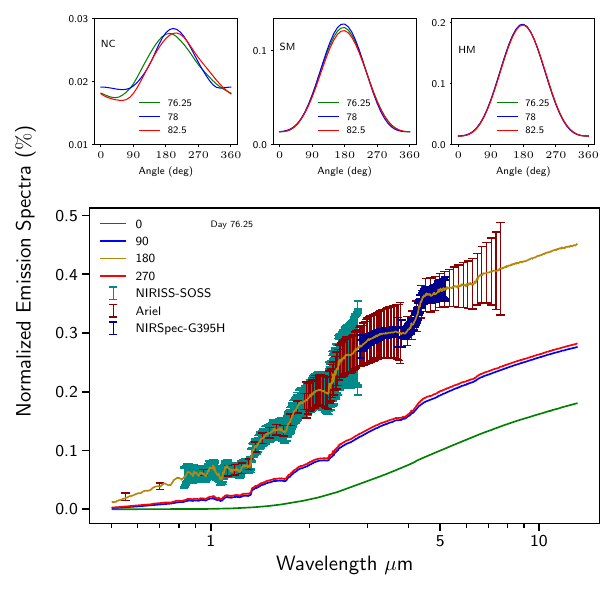}
    \caption{Emission flux normalized by the stellar flux, integrated 
    over a disk centered at the equator and different longitude angles $\varphi \in\{0, 90, 180, 270\}$~deg, at planet day $t = 76.25$. 
    A full phase-curve of the mean eclipse-depth is shown at the top 
    for three different days: $76.25$, $78.00$, and $82.50$ for the 
    three atmospheres (NC, SM, and HM). 
    Depending on the composition and chemistry of the atmosphere, 
    the variability in the flux (most visible for the NC atmosphere).
    The error bars for JWST and Ariel are included to illustrate the uncertainties for the two telescopes.}
    \label{fig:viewangle_flux}
\end{figure}

In Figure \ref{fig:viewangle_flux}, we present in main plot at 
bottom the emission flux spectrum (normalized by the stellar 
flux spectrum) at planet day $t = 76.25$; 
the flux is obtained by integrating over a disk centered at 
longitudes, $\varphi \in \{0,\, 90,\, 180,\, 270\}$~deg, all at 
latitude $\vartheta = 0$.
The disk center locations give the fluxes from the nightside, 
east terminator, dayside, and west terminator views of planet.
Here we show the spectra for the $Z = 12$ (HM) atmosphere, to 
demonstrate the $\varphi$-dependence of the spectrum even in the 
extreme low variability case.
The error bars for JWST and Ariel instruments are also indicated 
on the $\varphi = 180$~deg spectrum, to display the uncertainty 
on a spectrum that would be obtained;
the error bars are spread over the spectrum for clarity. 
The three subplots at the top of the figure show the 
$\lambda$-averaged black-body emission flux with the disk centered 
at $\vartheta = 0$ and over the full longitude range 
$\varphi \in [0, 360)$~deg, at three different days 
$t \in \{76.25,\, 78.00,\, 82.50\}$.
Here the $\lambda$-averaging is over the bandwidth
$[0.8, 2.8]~\mu$m, corresponding to the range 
for JWST/NIRISS-SOSS.

As can be seen in the bottom plot of the figure, the 
disk-integrated emission spectra obtained at different points on 
the planet are distinguishable.
The spectra are distinguishable by all three instruments longward 
of $\lambda \sim 1.3$~$\mu$m. Notice also the variability of the ``phase curves'' (i.e., 
band-averaged emission flux over $\varphi$) in time---particularly 
for the NC atmosphere (top row, left).
There, the collective movement of the hot and cold patches on the 
planetary scale near the $p =10^5$~Pa level is captured, as 
indicated by the movement of the peak disk-integrated emission 
longitude $\varphi_{\mbox{\tiny max}}(t) = (190, 200, 210)$~deg 
in time.
Animation of the temperature field during 
$t = [76.25,\, 82.50]$~days shows clearly the east and west 
terminator regions alternately become hotter and colder on a 
timescale of $\sim$3~days; see \citet{Changeatetal24} for the 
animation.
As for the SM and HM atmospheres (top row, middle and right, 
respectively), the peak emission longitude appears almost 
stationary near $\varphi = 180$~deg in the phase curves.
This is because, although the collective motion of the hot 
patches oscillate north--south about the substellar point, it 
does not vary much in the east--west direction, at the lower 
$p$-levels (see Figure \ref{fig:temp-map}, top row); recall 
that the $\lambda$-integrated flux is also reduced in these 
atmospheres (see Figure~\ref{fig:ar_jwst_flux}, left).

In the phase curves, the variation in peak amplitude longitude 
over the duration shown
$\Delta\varphi_{\mbox{\tiny max}}$ is $\sim$20~deg for the NC 
atmosphere, whereas 
$\Delta\varphi_{\mbox{\tiny max}} \lesssim 10$~deg for the SM and 
HM atmospheres.
In addition, the peak amplitude is shifted slightly eastward in 
the SM atmosphere, whereas there is no shift in the HM atmosphere; 
the latter is because the flux is washed out. 
Moreover, the variation in the {\it amplitude} of the peak flux 
for the NC, SM, and HM atmospheres are $\sim$3\%, $\sim$5\%, and $\sim$0.5\%, respectively; 
here the first value is slightly lower than the second because 
of the more efficient mixing of temperature at the $p \sim 10^5$~Pa 
level.
%and the $\Delta\lambda$ used in the averaging. 
The salient point, however, is that phase curves over multiple 
orbits can be used to help delineate the source of variability 
type (i.e., spatial, temporal, and spatiotemporal) as well as the 
metallicity in the atmosphere.

The phase offsets and amplitude variations in the phase curve, 
such as those shown in Figure \ref{fig:viewangle_flux}, are 
observable with both JWST and Ariel. 
Spectra can also be obtained in the observations, and for \WASPP\ 
have already been carried out with HST \citep{2017hst..prop15134E}, JWST/NIRISS \citep{2017jwst.prop.1201L},  
and JWST/NIRSpec-G395H \citep{2021jwst.prop.1729M}.
However, we note here that planets with atmospheres in different 
dynamical parameter regimes---e.g., hot-Jupiters with strong 
mechanical or thermal forcing in the deeper region 
($1~{\rm Pa} \lesssim p \lesssim 5~{\rm Pa}$) 
\citep{Choetal03,Skinetal23} as well as warm-Jupiters, sub-Neptunes, 
and super-Earths with weak dayside--nightside temperature contrasts 
or short rotation periods \citep[e.g.,][]{Choetal08,Kempton_2023, MeierValds_2023,SkinWei25}---will 
lead to different variability signatures than those 
presented here.

\section{Discussion}\label{sec:discussion}

In this paper, we have discussed the observable variability of 
medium-wave ($\sim$0.5~$\mu$m to $\sim$13~$\mu$m) flux for an 
ultra-hot-Jupiter \WASPP. 
We have utilized a high-resolution, retrieval-guided dynamics 
simulation to illustrate the time-varying signatures resulting 
from dynamic flow and temperature patterns on the planet. 
Reliable assessment of variability is obtained by post-processing 
the outputs from the accurate dynamics calculations with RT and 
chemistry. 
Simplified brightness flux calculations, as offered in many 
previous works, are also performed to compare with the more 
sophisticated (RT plus chemistry) treatment.

Our calculations show that the variability is highly 
dependent on the wavelength considered.
This is significant for observation of the same planet by 
different instruments such as JWST/NIRISS-SOSS, Ariel, and 
JWST/NIRSpec-G395H. 
The wavelength-integrated flux obtained with the instruments is 
also sensitive to the bulk metallicity and leads to the sensing 
of different pressure level regions (altitudes) of the atmosphere: 
$p \sim 10^5$~Pa level is probed in a H$_2$/He~(only) atmosphere 
while $p \sim 10^3$~Pa level is probed in a $Z \in \{1, 12\}$ 
metallicity atmosphere.
For this reason, the $Z \in \{1, 12\}$  atmospheres generally 
show reduced variability, as much of the outgoing radiation is 
absorbed before emerging. 
We have also discussed the observational constraints for JWST 
and Ariel in phase curves, which could provide important  
diagnostics of weather and climate patterns and help constrain 
atmospheric dynamics processes by mapping the motion of large 
atmospheric structures.

In addition, our calculations show that simple brightness 
temperature flux is sufficient for assessing the variability of 
\WASPP-like atmosphere, if the emergent level for the flux is 
known.
In another words, a full RT post-processing is not needed. 
This corroborates the brightness temperature flux approach taken 
in many past studies.
The dynamics along with the chemistry govern the spectral 
variability of the planet, through continuous 3D heating and 
cooling at different regions of the atmosphere. 
Modeling the 3D dynamics and active species distributions is
poorly constrained at present.
However, advancements can be made with more observations of the 
flux variability.

\section*{Acknowledgments}
The authors thank Jack W. Skinner for providing the dynamics 
simulation data. 
This publication is part of the project ``Interpreting exoplanet 
atmospheres with JWST'' with file number 2024.034 of the research 
programme ``Rekentijd nationale computersystemen'' funded in part 
by the Netherlands Organisation for Scientific Research (NWO) 
under grant \url{https://doi.org/10.61686/QXVQT85756}. 
This work used the Dutch national e-infrastructure with the support 
of the SURF Cooperative using grant no. 2024.034.

\bibliography{references}
\bibliographystyle{aasjournal}

\appendix
\section{Species Distribution}\label{sec:appendix1}
\begin{figure}[htb!]
    \includegraphics[width=0.45\linewidth]{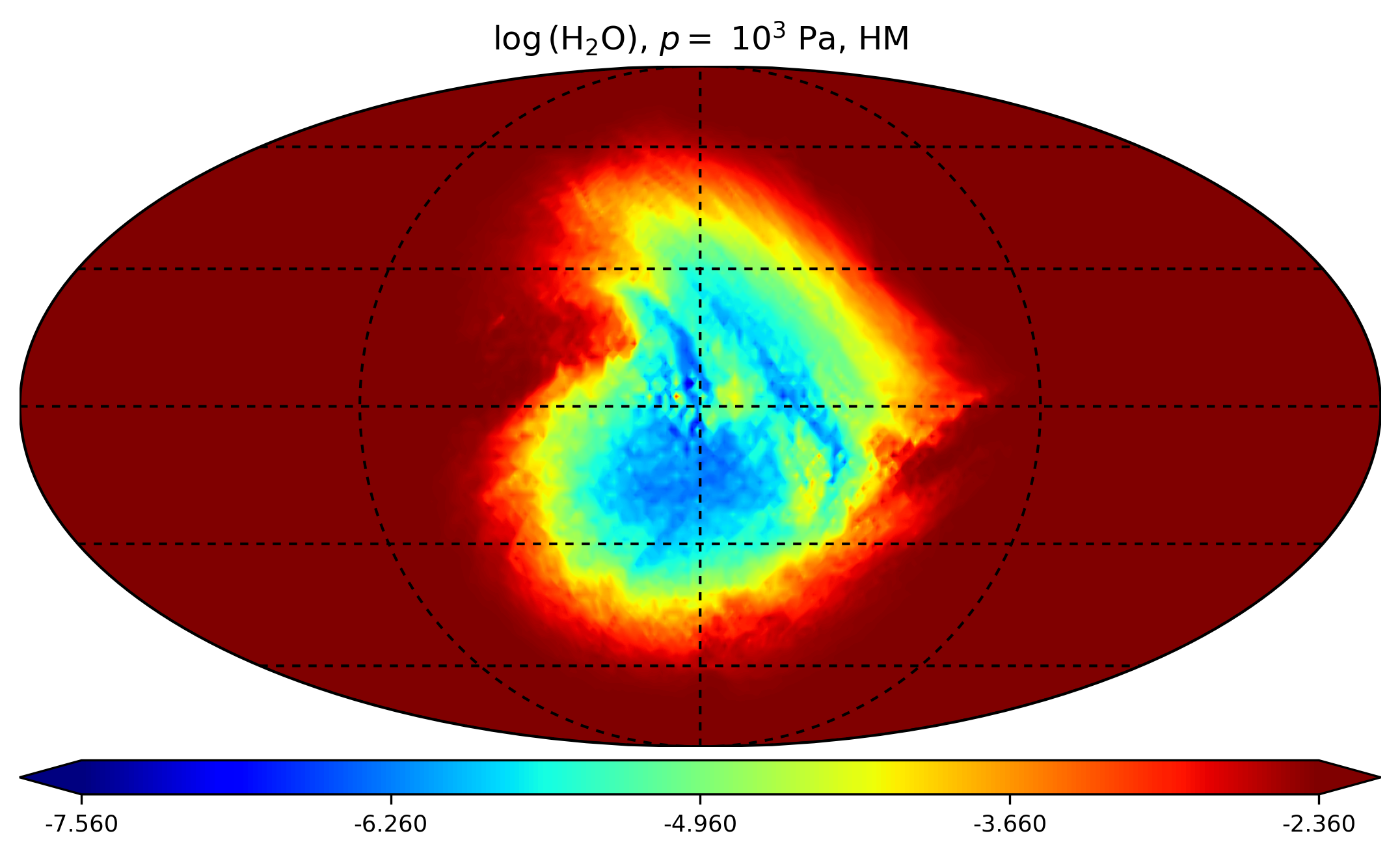}
    \hfill
    \includegraphics[width=0.45\linewidth]{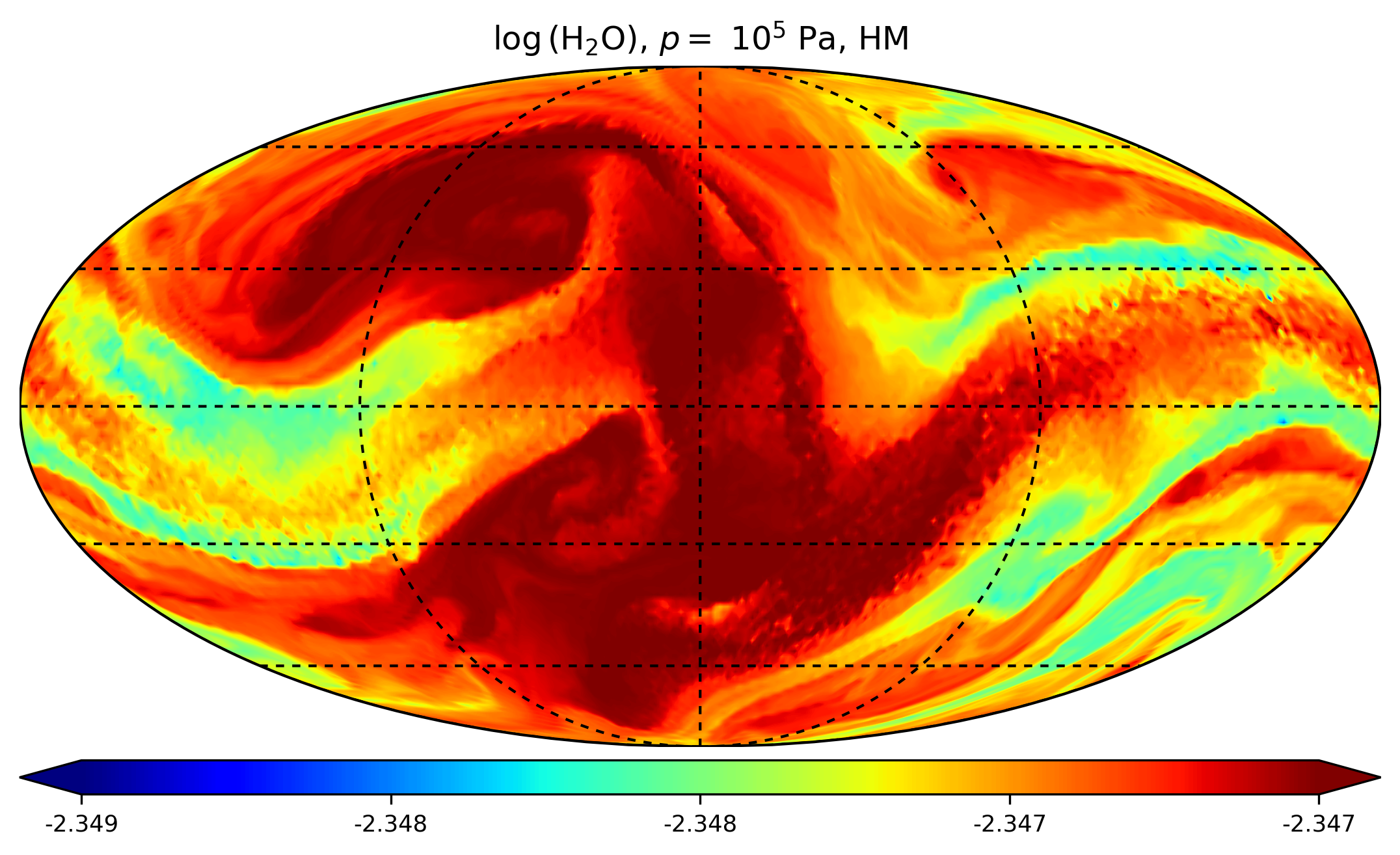}
    \includegraphics[width=0.45\linewidth]{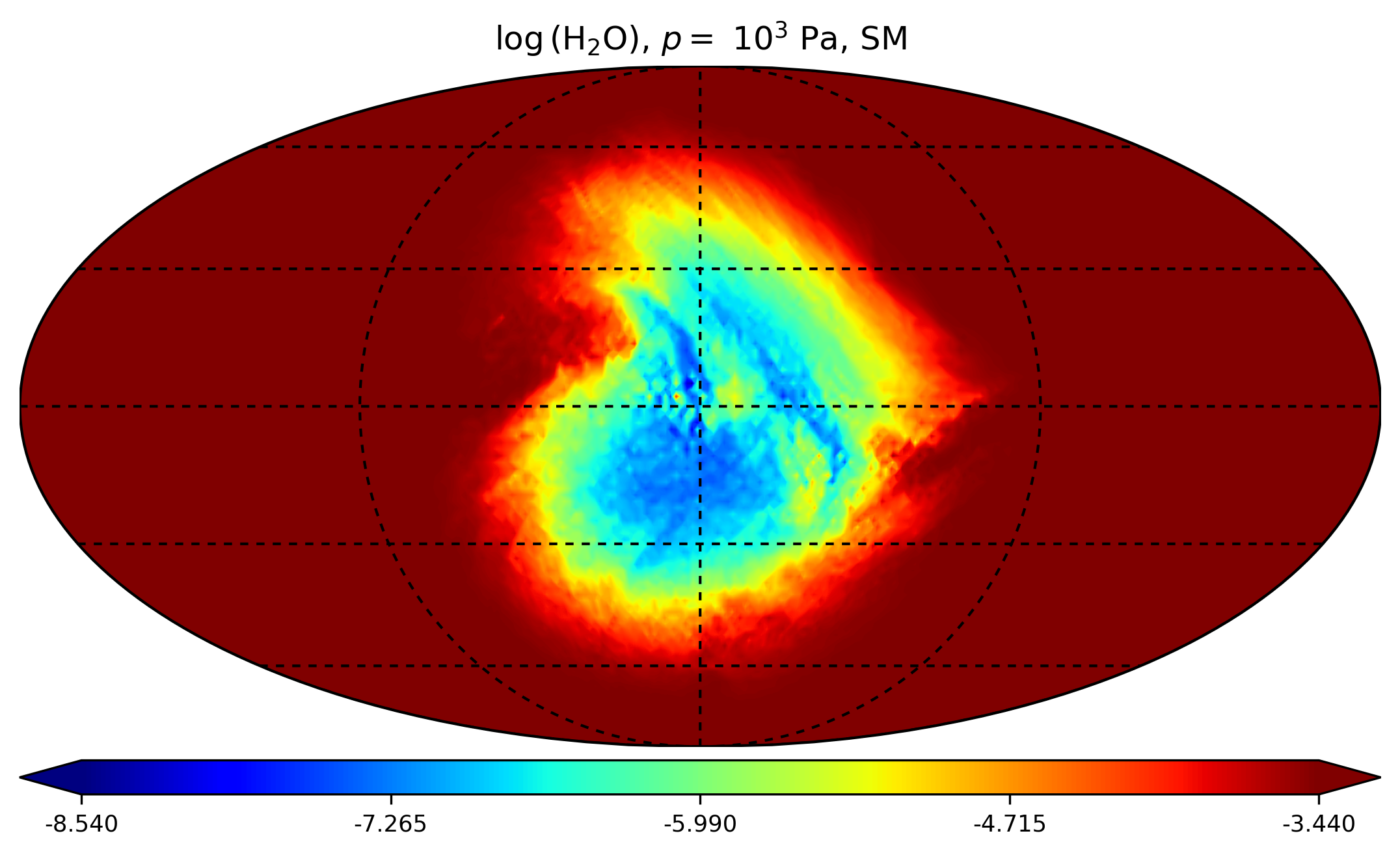}
    \hfill
    \includegraphics[width=0.45\linewidth]{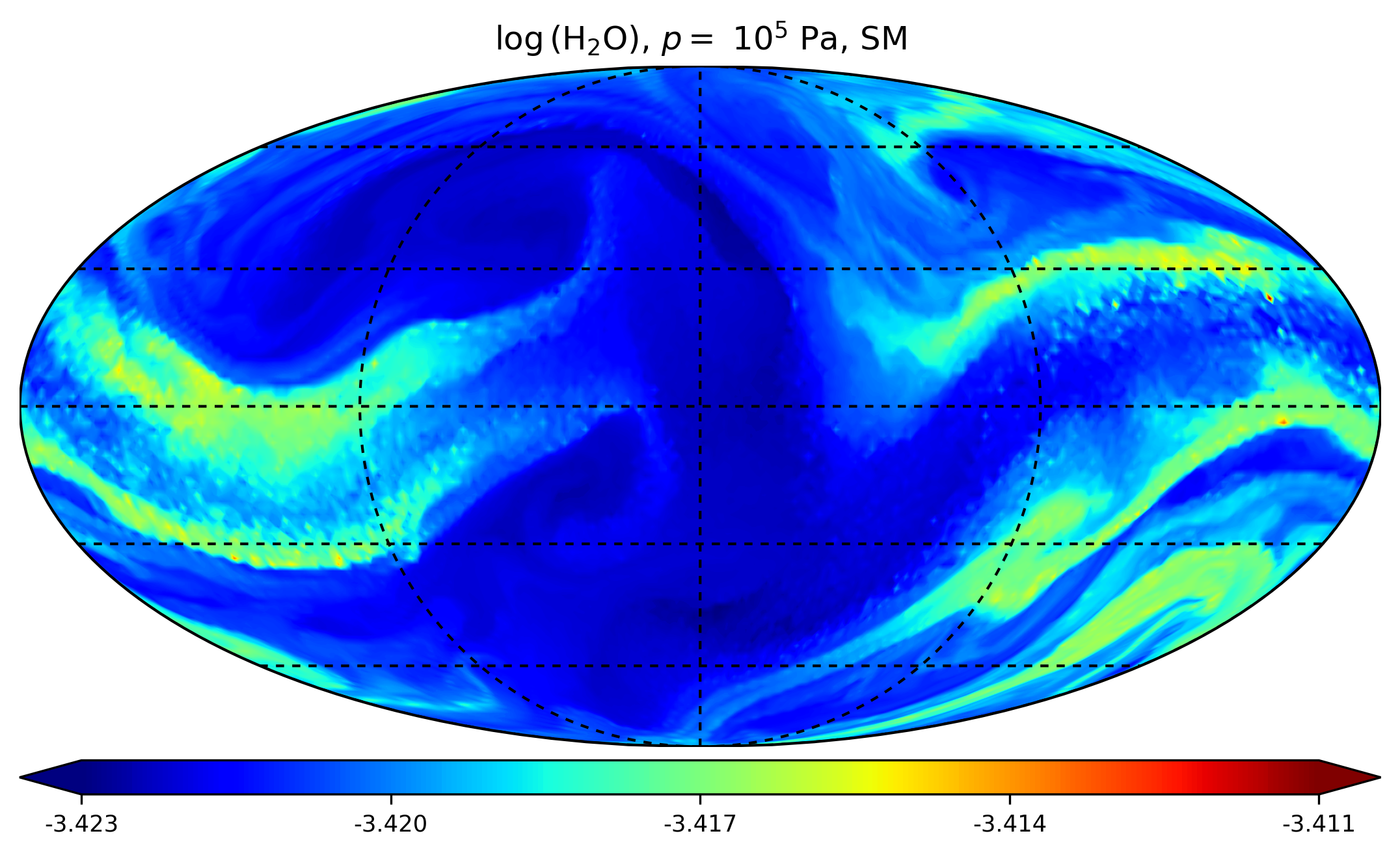}
    \includegraphics[width=0.45\linewidth]{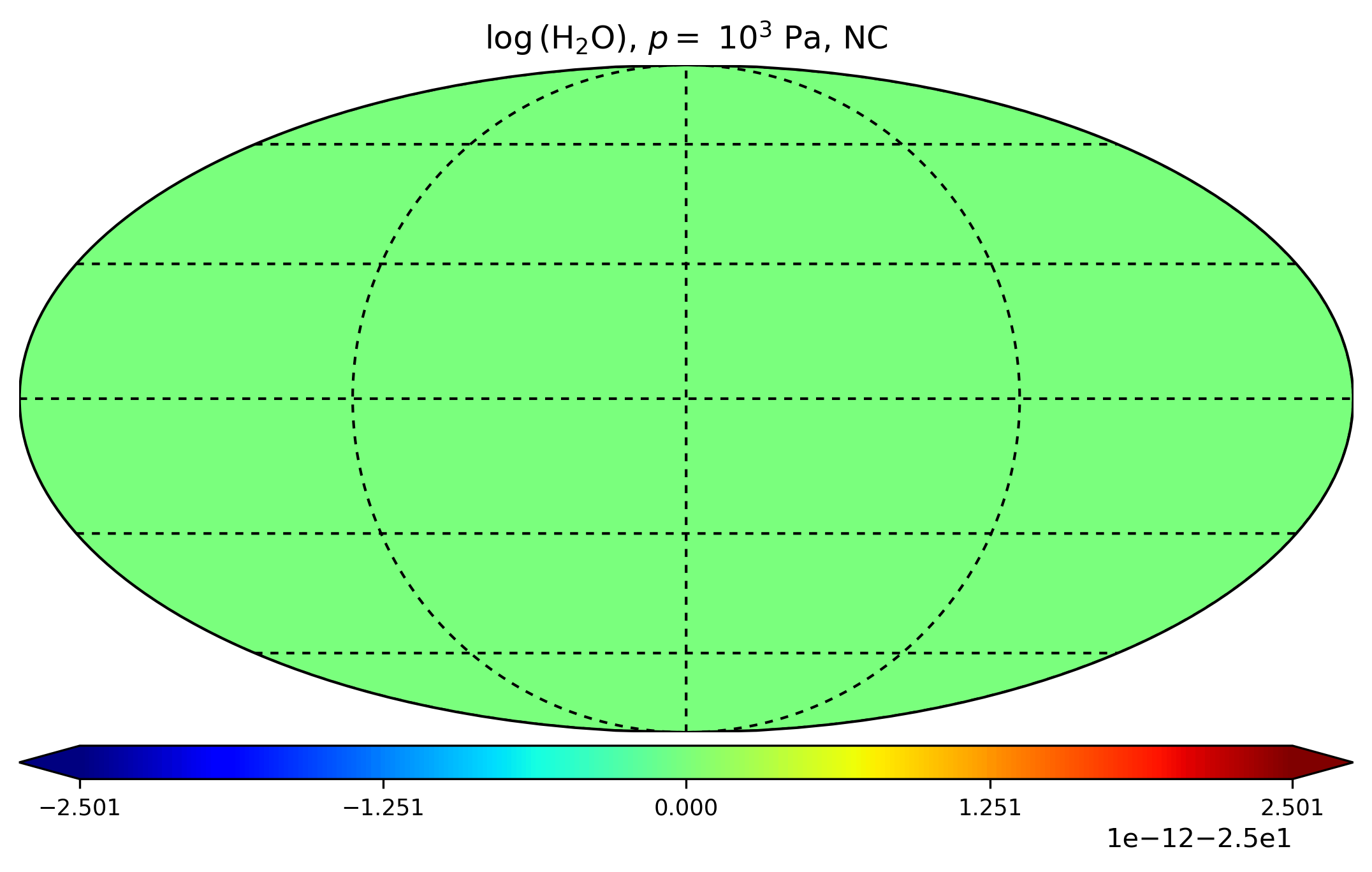}
    \hfill
    \includegraphics[width=0.45\linewidth]{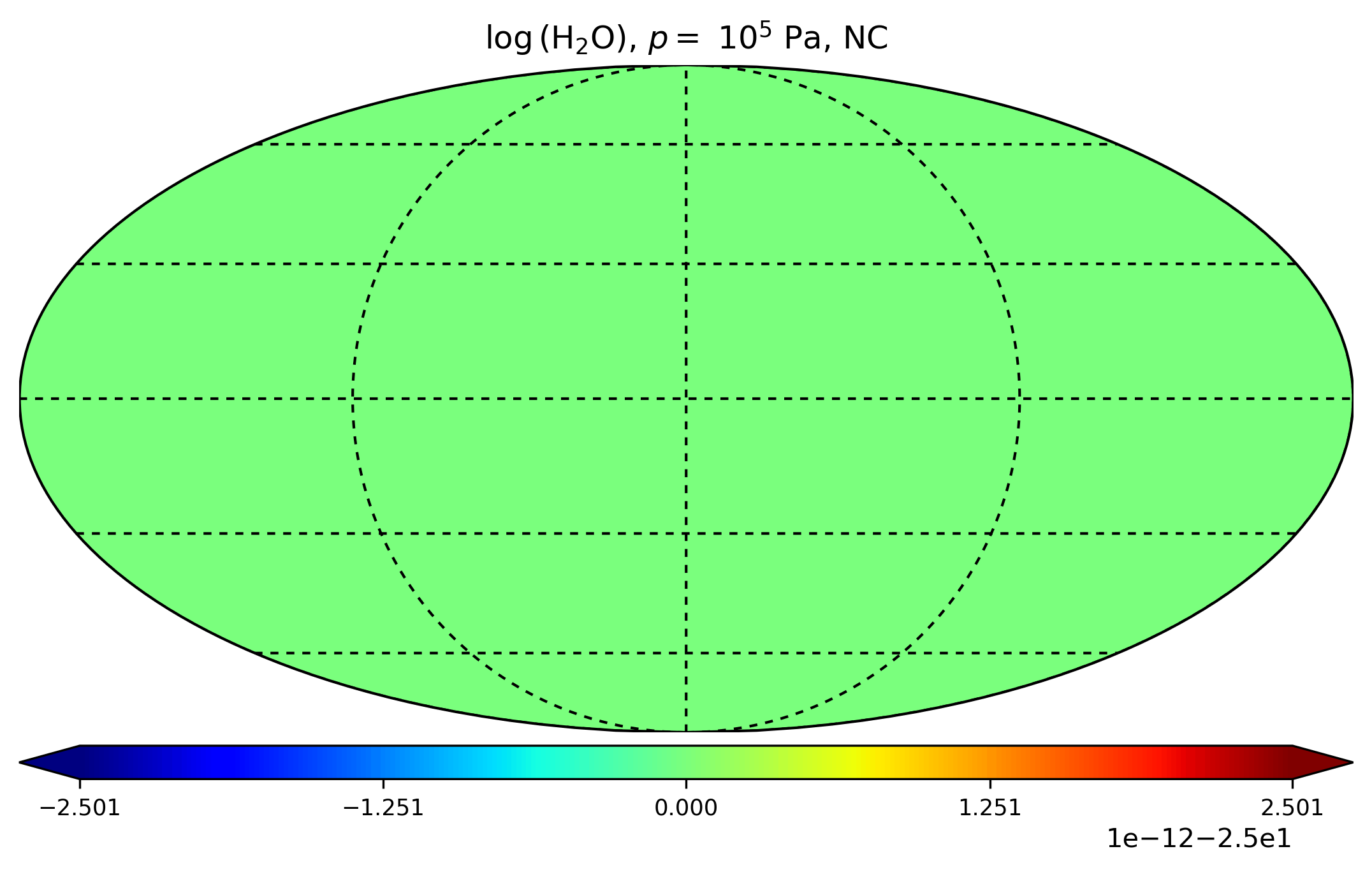}
    \caption{Example chemical species (mixing ratio) distribution 
    for H$_2$O at day $76$ at two pressure levels, 
    $p = 1\times\!10^3$~Pa (left column) and $p=1\times\!10^5$~Pa 
    (right column), and three types of atmospheres, HM with $Z =12$ 
    (top), SM with $Z = 1$ (middle), and NC with H$_2$--He only 
    (bottom).}
    \label{fig:chemistry}
\end{figure}

Figure \ref{fig:chemistry} shows the mixing ratio distribution of a key 
chemical species, H$_2$O, obtained from post-processing the temperature 
fields (see, e.g., Figure \ref{fig:temp-map}) with the GGChem code. 
The species distributions depend heavily on the underlying, dynamic 
temperature and the chemistry assumption: H$_2$--He (only), $Z = 1$, 
and $Z = 12$ atmospheres.

\newpage
\section{JWST/NIRSpec-G395H}
\begin{figure*}[htb!]
    \centering
    \includegraphics[width=0.49\textwidth,height=0.40\textwidth]{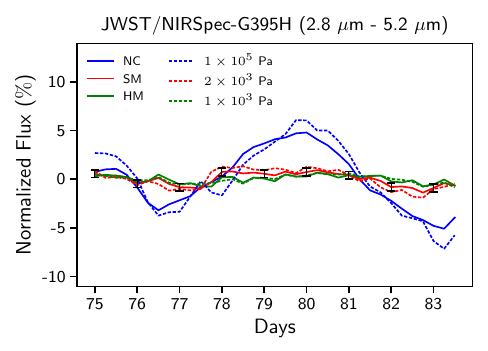}
    \caption{Normalized flux $\scrF$ in eclipse (i.e., dayside) 
    as a function of time for a \WASPP, integrated over the 
    wavelength ranges covered by JWST/NIRSpec-G395H.
    Full radiative transfer (RT) fluxes (full lines) and blackbody 
    brightness temperature fluxes (dashed lines) at the
    $p \in \{1 \times 10^5,\, 2 \times 10^3,\, 1 \times 10^3\}$~Pa 
    levels.
    Additionally, atmosphere with three different compositions (and 
    chemistry) are shown: 
    H$_2$/He atmosphere dominated by collision-induced absorption 
    (NC), solar metallicity atmosphere (SM), and super-solar 
    metallicity atmosphere (HM). 
    JWST/NIRSpec-G395H uncertainty for \WASPP\ dayside (secondary eclipse) 
    is also displayed and shows that the telescope would be able 
    to observe the variability of this planet, depending on the 
    metallicity. Note the variability amplitude is approximately 
    half of JWST/NIRISS-SOSS and Ariel (see 
    Figure~\ref{fig:ar_jwst_flux}). This is because of the 
    wavelength dependence of the variability -- JWST/NIRSpec-G395H 
    covers redder wavelengths ($> 2.8~\mu$m), where less 
    variability is seen (see Figure \ref{fig:flux_variation}).}
    \label{fig:jwst_nirspec}
\end{figure*}

Figure \ref{fig:jwst_nirspec} shows the variability for 
JWST/NIRSpec-G395H.
This should be compared with the variability for JWST/NIRISS-SOSS
and Ariel in Figure \ref{fig:ar_jwst_flux}.
The latter two cover different wavelength ranges.
\label{appendix2}

\end{document}

%% file: chocros.tex
\usepackage{amsmath}
\usepackage{amssymb}
\usepackage{upgreek}
\usepackage{bbm}
\usepackage{dsfont}
\usepackage{sansmath}
\usepackage{mathrsfs}
\usepackage{yfonts}
\usepackage{euscript}

\def\d{{\rm d}}
\def\D{{\rm D}}

\def\R{{\mathbbm{R}}}

\def\calF{{\cal F}}

\def\scrF{\mathscr{F}}

\def\gsim{\;\rlap{\lower 2.5pt\hbox{$\sim$}}\raise 1.5pt\hbox{$>$}\;}
\def\lsim{\;\rlap{\lower 2.5pt\hbox{$\sim$}}\raise 1.5pt\hbox{$<$}\;}

\def\beq{\begin{equation}}
\def\eeq{\end{equation}}

\newcommand{\WASPP}{WASP-121\,b}

\usepackage{color}
\definecolor{com_red}{rgb}{.8,.2,0.1}
\definecolor{com_burg}{rgb}{.45,.05,0.15}
\definecolor{com_turq}{rgb}{.00,.35,0.65}
\definecolor{ins_green}{rgb}{.2,.5,0.2}